\documentclass[12pt]{article}
\usepackage{graphicx}
\usepackage{amssymb,epsf}
\def\bseq{\begin{subequation}}  
\def\eseq{\end{subequation}}
\def\bsea{\begin{subeqnarray}}  
\def\esea{\end{subeqnarray}}

\newcommand{\bbox}{\lower.2ex\hbox{$\Box$}}

\newcommand{\beq}{\begin{equation}}
\newcommand{\eeq}{\end{equation}}
\newcommand{\bea}{\begin{eqnarray}}
\newcommand{\eea}{\end{eqnarray}}
\newcommand{\ena}{\end{eqnarray}}

\newcommand {\non}{\nonumber}

\renewcommand{\a}{\alpha}
\renewcommand{\b}{\beta}

\renewcommand{\d}{\delta}

\newcommand{\pa}{\partial}
\newcommand{\g}{\gamma}
\newcommand{\G}{\Gamma}

\newcommand{\e}{\epsilon}
\newcommand{\z}{\zeta}

\renewcommand{\l}{\lambda}

\newcommand{\m}{\mu}

\newcommand{\p}{\pi}

\newcommand{\Db}{\bar{D}}

\newcommand{\Phib}{\bar{\Phi}}

\newcommand{\ad}{{\dot{\alpha}}}

\newcommand{\Del}{\nabla}

\begin{document}
\begin{titlepage}
\begin{flushright}
IFUM--892--FT \\
Bicocca-FT-06-14\\

\end{flushright}

\vspace{.4cm}

\noindent{\Large \bf   Real versus complex $ \b$--deformation  of the $  {\cal N}=4$\\ 
\noindent planar super Yang-Mills theory }
 
 \vspace{0.4cm}

\vfill
\begin{center}

{\large \bf Federico Elmetti$^1$,~Andrea Mauri$^1$,~Silvia
Penati$^2$, \\

\vspace{0.3cm} 
Alberto Santambrogio$^1$ and
Daniela Zanon$^1$}\\

\vspace{0.5cm}

{\small $^1$ Dipartimento di Fisica, Universit\`a di Milano and\\
INFN, Sezione di Milano, Via Celoria 16, I-20133 Milano, Italy\\

\vspace{0.1cm}
$^2$ Dipartimento di Fisica, Universit\`a di Milano--Bicocca and\\
INFN, Sezione di Milano--Bicocca, Piazza della
Scienza 3, I-20126 Milano, Italy}\\

\end{center}
\vfill
\begin{center}
{\bf Abstract}
\end{center}
{This is a sequel of our paper hep-th/0606125 in which we have studied 
 the ${\cal N}=1$ $SU(N)$ SYM theory obtained as a
marginal deformation of the ${\cal N}=4$ theory, with a complex
deformation parameter $\b$ and in the planar limit. There we have addressed 
the issue of conformal invariance imposing the theory 
to be finite and we have found that finiteness requires reality of the 
deformation parameter $\b$.  

\noindent In this paper we relax  the finiteness request and look for a theory  that in the planar limit has vanishing beta 
functions. We perform explicit  calculations up to five loop order: we
find that the conditions of  beta function vanishing can be achieved with a complex deformation parameter, 
but the theory is not finite and the result depends on the arbitrary choice of the subtraction procedure.
Therefore, while the finiteness condition leads to a scheme independent result, so that the conformal invariant theory with a real 
deformation is physically well defined, the condition of vanishing beta function leads to a result which is scheme dependent 
and therefore of unclear significance. 

\noindent In order to show that these findings are not an artefact of dimensional regularization, 
we  confirm our results within the differential renormalization approach.} 
\vspace{2mm} \vfill \hrule width 3.cm
\begin{flushleft}
e-mail: federico.elmetti@mi.infn.it\\
e-mail: andrea.mauri@mi.infn.it\\
e-mail: silvia.penati@mib.infn.it\\
e-mail: alberto.santambrogio@mi.infn.it\\
e-mail: daniela.zanon@mi.infn.it
\end{flushleft}
\end{titlepage}

\section{Introduction}

Recently we have studied marginal deformations of the ${\cal N}=4$ supersymmetric Yang-Mills theory best known 
as $\beta$--deformations. 
These theories are obtained through the following modification of the ${\cal N}=4$ theory: one enlarges the space 
of parameters adding 
to the gauge coupling $g$ two complex couplings $h$ and $\b$. These new parameters enter  the chiral superpotential 
via the substitution
\beq 
ig~{\rm Tr}(  ~\Phi_1
\Phi_2 \Phi_3 - ~ \Phi_1 \Phi_3 \Phi_2 ~)~\longrightarrow ~ih~{\rm
Tr}\left( ~e^{i\pi\b} ~\Phi_1 \Phi_2 \Phi_3 - e^{-i\pi\b}~ \Phi_1
\Phi_3 \Phi_2 ~\right) 
\label{deformation} 
\eeq 
It has been argued that these $\b$-deformed ${\cal N}=1$ theories become conformally
invariant if the constants $g$, $h$ and $\b$ satisfy one equation in the space of parameters  \cite{LS}. Of  
course it is of interest to find this condition explicitly.
For the case
of $\b$ real and in the planar limit we have shown \cite{MPSZ} that to all perturbative orders  this equation is 
simply given by
\beq h\bar{h}=g^2 
\label{betareal} 
\eeq 
The corresponding conformal theory represents
the exact field theory dual to the Lunin--Maldacena supergravity 
background \cite{LM}. Further confirmation of this result can be found in \cite{Khoze, kovacs}.

In a recent paper \cite{EMPSZ} we have extended our study to the case of 
complex
$\b$ \cite{RSS2}. The analysis was performed in the planar limit, using a perturbative approach, superspace techniques and 
dimensional regularization. 
With the aim of addressing the issue of conformal invariance we have investigated the finiteness of the theory. 
In fact simply imposing the finiteness of the two-point chiral correlators we  found that  only {\em real} values of the 
parameter $\b$ are allowed, thus leading to the condition  in (\ref{betareal}).  Being the theory finite,
this result  is obviously independent of the renormalization scheme adopted throughout the calculation.
The corresponding theory is conformal invariant and perfectly well defined.

On one hand this result might be somewhat surprising since the expectation was to find an equation for the 
parameters, $g$ real and 
$h$ and $\b$ complex, with no additional constraints. On the other hand the request of real $\b$ seems to be 
in agreement with results 
in the string dual approach where singularities appear in the deformed metric as soon as $\b$ acquires an 
imaginary part \cite{LM,GN,KT}. Our findings are also consistent with results 
concerning the integrability
of the theory \cite{FRT,BCR}.

In this paper we reexamine the problem imposing less restrictive requirements. 
Here in order to have a 
conformal theory we simply ask the gauge beta function and the chiral beta function  to vanish.
The general 
strategy we have in mind is  to define the theory at its conformal point looking for a surface of renormalization group
fixed points in the space of the coupling constants. This amounts to perform a coupling constant reduction
by expressing the chiral couplings as functions of the gauge coupling $g$. This operation has an immediate 
consequence:  we are forced to work perturbatively in powers of $g$ instead of powers of loops. This 
allows different loop orders to mix and in general the conditions which insure finiteness become different from the
conditions for vanishing beta functions. Therefore standard finiteness theorems \cite{PW,GMZ} for the 
gauge beta function cannot be used. 

We perform explicit  calculations up to five loops and find that the condition of vanishing beta functions can be accomplished with 
a complex deformation parameter, but the theory is not finite.  Thus we are forced to renormalize the theory and consequently the result is dependent on the arbitrary  choice of the subtraction procedure. Of course if we want to recover 
a result that does not depend on the renormalization scheme we have to impose finiteness and then we go back to a real 
deformation parameter.

In order to make sure that our findings are independent of the regularization procedure we have adopted, i.e. 
dimensional regularization 
\footnote{From our experience dimensional regularization always works but it is often questioned.},
we have redone various calculations within the differential renormalization approach and  confirmed the results.

It is worthwhile emphasizing that the five-loop calculation of the planar gauge beta function is a highly non trivial exercise. 
We have accomplished it through the use of improved superspace techniques \cite{GZ,GMZ} in conjunction with a lot of  
ingenuity in the D-algebra manipulations. Our result gives indication that a generalization of the standard finiteness theorems \cite{PW,GMZ} for the 
gauge beta function holds, i.e. if the matter chiral beta function vanishes up to ${\cal{O}}(g^n)$ then the gauge beta function is guaranteed to vanish up to ${\cal{O}}(g^{n+2})$.

The paper is organized as follows. In Section 2 we  describe the general approach and briefly review our previous calculation 
\cite{EMPSZ}. 
In Section 3 we present  the evaluation of the chiral and vector beta functions. We explicitly show how the conditions of vanishing beta functions do not give a finite theory and explain how the dependence on the renormalization scheme adopted affects the result. In Section 4 we discuss the calculation within 
the differential 
renormalization approach with the use of analytic regularization. Final comments are in our conclusions.

\section{The general setting and a brief review of  conformal invariance of the $\b$--deformed theory via finiteness}

We consider the ${\cal N}=1$ $\b$--deformed  action written in terms of the
superfield strength $W_\a= i\Db^2(e^{-gV}D_\a e^{gV})$ and  three chiral superfields 
$\Phi_i$ with $i=1,2,3$.
With notations as in \cite{superspace} we have
\bea 
S &=&\int d^8z~ {\rm
Tr}\left(e^{-gV} \Phib_i e^{gV} \Phi^i\right)+ \frac{1}{2g^2}
\int d^6z~ {\rm Tr} (W^\a W_\a)\nonumber\\
&&+ih  \int d^6z~ {\rm Tr}( ~q ~\Phi_1 \Phi_2 \Phi_3 - \frac{1}{q}~
\Phi_1 \Phi_3 \Phi_2 ~)
\nonumber \\
&& + i\bar{h}\int d^6\bar{z}~ {\rm Tr} ( ~\frac{1}{\bar{q}}~ \Phib_1
\Phib_2 \Phib_3 - \bar{q} ~\Phib_1 \Phib_3 \Phib_2~ )\qquad,\qquad
q\equiv e^{i\pi\b}  \label{actionYM} 
\eea 
where $h$ and $\b$ are
complex couplings and $g$ is the real gauge coupling constant. 
In the undeformed  ${\cal N}=4$ SYM theory one has $h=g$ and $q=1$.
In the present case it is convenient to define
\beq
h_1\equiv h q\qquad\qquad h_2\equiv \frac{h}{q} 
\eeq
and work with couplings $g$, $h_1$ and $h_2$.

In the spirit of \cite{LS} (see also \cite{oehme}-\cite{JZ}) the idea is to find a surface of
renormalization group fixed points in the space of the coupling constants. 
To this end one reparametrizes these couplings in terms of the gauge coupling $g$.
In fact, since in the planar limit for each diagram the color factors 
from chiral vertices is always in terms of  the products $h_1^2\equiv
h_1\bar{h}_1$ and $h_2^2\equiv h_2\bar{h}_2$, we express directly
$h_1^2$ and  $h_2^2$ as power series in the coupling $g^2$ as
follows 
\bea
&& h_1^2=a_1 g^2+a_2 g^4+a_3 g^6+\dots \nonumber\\
&&h_2^2=b_1 g^2+b_2 g^4+b_3 g^6+\dots 
\label{expansion} 
\eea

The final goal is to study the condition that in the large $N$ limit the couplings have to satisfy
in order to guarantee the conformal invariance of the theory for
complex values of $h$ and $\b$. 

In the large $N$ limit for real values of $\b$, i.e. if 
 $q\bar{q}=1$,  the $\b$-deformed theory
becomes exactly conformally invariant if the condition (\ref{betareal})
is satisfied \cite{MPSZ}. 
In this case the chiral couplings differ only
by a phase from the ones of the ${\cal N}=4$ SYM theory and  the
planar limits of the two theories are essentially the same.

When $q\bar{q}\neq 1$, in order to isolate the relevant terms and drastically simplify the analysis, it is convenient 
\cite{RSS} to study the condition of
conformal invariance considering the difference between contributions computed in the
 $\b$-deformed theory and the corresponding ones in
the ${\cal N}=4$ SYM theory (which is finite and with vanishing beta function). The simplification is  due essentially to the following facts:  when computing the difference between graphs  
in the $\b$-deformed and in the ${\cal N}=4$ theory 
we need not consider diagrams that contain only gauge-type
vertices since their contributions is the same in the two
theories.  Instead we concentrate on divergent graphs that
contain either only chiral vertices or mixed chiral and gauge
vertices. In fact the relevant terms come from the chiral vertices that are actually different in the two theories. 
Addition of vectors simply modifies the color due to the chiral
vertices by the multiplication of $g^2$ factors which are the same for both theories. 

The idea is to proceed perturbatively in superspace. The propagators for the
vector and chiral superfields, and the interaction vertices are
obtained directly from the action in (\ref{actionYM}). Supergraphs
are evaluated performing the $D$-algebra in the loops and the
corresponding divergent integrals are computed using dimensional
regularization in $n=4-2\e$.
\vspace{0.3cm}

In \cite{EMPSZ} these techniques were used to impose the condition of finiteness on the $\b$-deformed theory 
and to this end it was sufficient to require finiteness of the two-point 
chiral correlator.
We review the relevant steps of the calculation 
performed in \cite{EMPSZ} and refer the reader to that paper for technical details.

At order $g^2$ we have to consider one-loop divergent diagrams in the $\b$-deformed 
 and in the ${\cal N}=4$ theory and compute the difference. This amounts to the 
evaluation of chiral bubbles and gives the following divergent contribution 
to the chiral propagator
\beq
\frac{N}{(4\p)^2}~\left[
h_1^2+h_2^2-2g^2\right]~ \frac{1}{\e} 
\label{leftover1loop} 
\eeq

Using the expansions in
(\ref{expansion}), in order to obtain a finite result we have to impose the condition
\beq 
{\cal O}(g^2): \qquad\qquad \qquad  a_1+b_1-2=0 
\label{order1} 
\eeq 
In fact  we have shown \cite{EMPSZ} that the condition 
\beq
h_1^2 + h_2^2 = 2g^2
\label{finite} 
\eeq
ensures conformal invariance
up to three loops in the planar limit.  For the chiral two-point function the
only divergences come from the one-loop bubble  and this implies that 
up to order $g^6$, we find the following additional requirements 
\bea
&& {\cal O}(g^4): \qquad\qquad \qquad a_2+b_2=0 
\nonumber\\
&&{\cal O}(g^6): \qquad\qquad \qquad 
a_3+b_3=0  
\eea 
When we move up to four loops we can repeat the same reasoning as above. 
Indeed using the condition in (\ref{finite}) 
we can show that all the four--loop 
diagrams that either contain vector lines on chiral bubbles or
consist of various arrangements of chiral bubbles are not relevant. 
We simply need to focus on a new type of chiral divergent structure, 
the one shown in Fig.1.\\

\begin{figure} [ht]
\begin{center}
\epsfysize=4cm\epsfbox{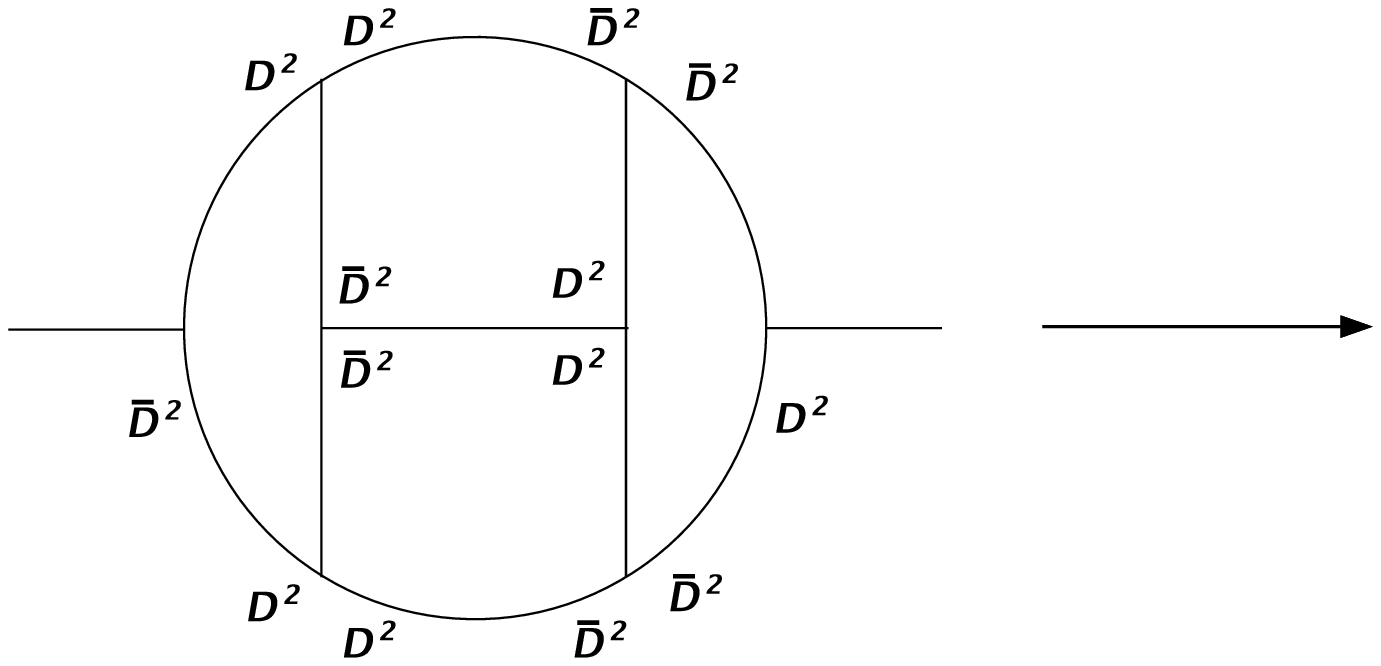} \hspace{0,4cm} \epsfysize=4cm\epsfbox{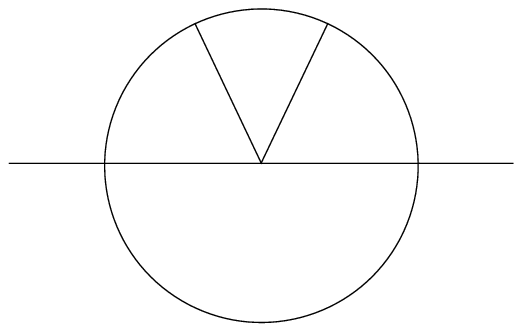}
\caption{Four-loop supergraph and its associated relevant bosonic integral}
\end{center}
\end{figure}

From the four-loop calculation in \cite{EMPSZ} we find that, computing the difference with the corresponding 
contribution from the ${\cal N}=4$ theory and using the
expansions  (\ref{expansion})  finiteness is achieved if
\beq
{\cal O}(g^8): \qquad \quad \quad
a_4+b_4-\frac{5}{2}~\z(5)~N^3~\frac{1}{(4\p)^6}(a_1-b_1)^4=0
\label{g8} 
\eeq 
For later convenience we define 
\beq 
A \equiv \frac{N}{(4\pi)^2} (a_4 + b_4)  \qquad \qquad
B \equiv - \frac{5}{2} \z(5) \frac{N^4}{(4\p)^8}  (a_1 - b_1)^4
\label{AB}
\eeq
so that the previous condition becomes 
\beq
A+B=0
\label{cond1}
\eeq
Then we  move to the next order. 
If we were following a standard procedure, i.e. canceling divergences order by
order in loops, having canceled the $1/\e$ pole terms at lower orders we would be guaranteed of the vanishing 
of the $1/\e^2$ terms at the next order.
In our case, instead, we have imposed the finiteness condition order by
order in $g^2$. At the order $g^8$ this has led us to the
relation (\ref{cond1}) which allowed us to cancel the $1/\e$ pole from the
one-loop diagram with the $1/\e$ pole from the four-loop diagram.
When computing the chiral two-point function, 
these one- and four-loop structures show up at order $g^{10}$ as
subdivergences in two-loop and five-loop integrals respectively.
It is easy to realize that they  produce $1/\e^2$-pole terms. In \cite{EMPSZ} we have shown that in order 
to cancel the $1/\e^2$ terms one has to impose $A=B=0$, i.e.
\beq
a_1 = b_1 = 1 \qquad {\rm and } \qquad a_4 + b_4 = 0
\label{first} 
\eeq
We note that at the order $g^8$ the finiteness condition (\ref{cond1}) is not sufficient to insure
the vanishing of the chiral beta function which turns out to be proportional to $A+4B$ 
(see eq.(\ref{gamma8}) in the next Section). Therefore at this order the theory is finite {\em but} the beta functions do not vanish. 
However if we take into account the finiteness condition from 
the order $g^{10}$ we end with $A=B=0$ which leads to a  theory  finite and 
at a RG fixed point. 
  
Under the conditions in (\ref{first}) $1/\e$ divergences at five and two loops are 
automatically canceled. Thus at order $g^{10}$ the only divergence in the chiral propagator
comes again from the one-loop bubble eq.(\ref{leftover1loop}) and we are forced to impose 
\beq 
a_5+b_5=0 
\label{g10}
\eeq
In \cite{EMPSZ} we have shown  that new chiral graphs at six loops and higher
are not relevant. Therefore, everything is controlled by the cancellation of $1/\e$ divergences
at one and four loops and of $1/\e^2$ poles at two and five loops. These patterns repeat themselves
at the order $(g^2)^{4k}$ and at the order $(g^2)^{4k+1}$ respectively.

The final solution is simply (see \cite{EMPSZ} for details)
\beq a_1=b_1=1 \qquad\qquad
a_n=b_n=0~~~~~n=2,3,\dots
\label{finalcoeff}
\eeq
which implies $\sinh(2\pi{\rm Im}\b) \sim (h_1^2 - h_2^2) =0$. Therefore, the $\b$-deformed SYM theory is 
finite only for $\b$ real and, as already stressed, the beta functions also vanish.

We emphasize that this result is independent of the renormalization scheme: had we done the calculation using a different scheme 
the condition of finiteness would have led us to the solution $\b$ real.

In the next section we will relax the finiteness requirement. We want to find the condition that the couplings have to satisfy 
in the large $N$ limit
in order to guarantee the vanishing of the chiral and gauge beta functions. We will find that in this case complex values of 
$\b$ are allowed but the resulting conformal invariant theory depends on the renormalization scheme.

\section{Conformal invariance of the $\b$--deformed theory via vanishing of the chiral and gauge beta functions}

Now we go back to the action in (\ref{actionYM}) and compute perturbatively in the large $N$ limit the chiral and gauge 
beta functions. The request of vanishing beta functions will identify a conformal field theory. 

\noindent First we consider the chiral 
beta function $\b_h$. It is well-known that in minimal subtraction scheme $\b_h$ is proportional to the anomalous dimension $\g$ of the elementary fields and the condition $\b_h=0$ can be conveniently traded with $\g=0$.  In our case, even working in a generic scheme, one can easily convince oneself that at a given order in $g^2$ the proportionality relation between $\b_h$ and $\g$ gets 
affected only by terms proportional to lower order contributions to $\g$. Therefore, if we set
$\g=0$ order by order in the coupling, we are guaranteed that $\b_h$ vanishes as well. 

Thus we impose the 
vanishing of $\g$ which we obtain from the computation of the two-point chiral correlator. Up to three loops nothing 
new happens: the condition in (\ref{finite}) insures the vanishing of $\g$ up to the order $g^6$ and correspondingly also  
$\b_h$ is zero.  Moreover up to this order we can use the results in \cite{GMZ} and we are guaranteed that also 
the gauge beta function $\b_g$ is zero up to the order $g^9$. This is easily understood since in spite of the redefinition 
in (\ref{expansion}) the request of vanishing anomalous dimensions up to order $g^6$ works order by order in the loop expansion 
so that general finiteness theorems \cite{PW,GMZ} hold. At this stage the coefficients in (\ref{expansion}) have to satisfy
\bea
&&{\cal O}(g^2): \qquad\qquad \qquad  a_1+b_1-2=0 \nonumber\\
&& {\cal O}(g^4): \qquad\qquad \qquad a_2+b_2=0 
\nonumber\\
&&{\cal O}(g^6): \qquad\qquad \qquad 
a_3+b_3=0 \label{order2-3} 
\eea 

Things become more subtle at  ${\cal{O}}(g^8)$: here the only way to achieve the vanishing of $\g$ is to mix contributions 
from one loop with contributions from four loops.
Repeating the calculation of the divergent integrals,
the result is proportional to 
\beq
\frac{1}{\e}\left[ A\left( \frac{\m^2}{p^2}\right)^{\e} + B\left( \frac{\m^2}{p^2}\right)^{4\e}\right]
\label{forgamma}
\eeq
where $A$ and $B$ were defined in (\ref{AB}) and we have explicitly indicated
the factors coming from dimensionally regulated integrals at one and
four loops (here $p$ is the external momentum and $\m$ is the standard 
renormalization mass). The anomalous dimension is given directly by the 
finite $log$ term in (\ref{forgamma})
and then we see that at order $g^8$ the vanishing of the anomalous dimension 
$\g$ requires
\beq
{\cal O}(g^8): \qquad\qquad \qquad A+4B=0
\label{gamma8}
\eeq
We emphasize that at this order this condition ensures the vanishing 
of $\g$ and $\b_h$ , but as it appears in (\ref{forgamma}) the theory is {\it not finite}.  We will come back to this point and discuss its implications below. First we want to show 
that the condition in (\ref{gamma8})  is sufficient to insure that 
$\b_g$ is zero up to the order $g^{11}$.

 Contributions to the gauge beta function at ${\cal{O}}(g^{11})$ come from two- and five-loop diagrams. Using standard 
superspace methods the two-loop calculation is straightforward, but  at five loops the number of diagrams involved is 
large and the calculation looks rather repulsive.\footnote{We recall that in \cite{JJN} a calculation of similar difficulty 
was attempted: the four-loop gauge beta function including nonplanar graphs. In that case the relevant coefficient was obtained by 
an indirect assumption because a direct calculation was too involved.}

In fact using the background field method and covariant supergraph techniques we are able to perform this high loop 
calculation exactly. 
We take advantage of the results obtained in \cite{GMZ} where the structure of higher-loop ultraviolet divergences in 
SYM theories was analyzed using the superspace background field method and supergraph  covariant D-algebra  \cite{GZ}. 
Using this approach   contributions to the effective action  beyond one loop can be written in terms of the vector 
connection $\G_a$ and the field strengths $W_\a$, $\bar{W}_\ad$, but not of the spinor connection $\G_\a$. This result 
allows to draw strong conclusions on the structure of UV divergences in SYM theories.
It was shown \cite{GMZ} that in regularization by dimensional reduction UV divergent terms can be obtained by computing a 
special subset of all possible supergraphs.
The reasoning can be summarized as follows: at any loop order (with the exception of one loop), after subtraction of UV and 
IR divergences, the infinite part of contributions to the effective action is local and gauge invariant. By superspace 
power counting and gauge invariance it must have the form
\beq
\G_{\infty}=z(\e)~{ \rm{Tr}} \int d^4x~d^4\theta~\G^a\G_b(\d_a^{~b}-\hat{\d}_a^{~b})
\label{covstructure}
\eeq
where $\G^a$ is the vector connection from the expansion of the covariant derivatives, i.e. $\Del_a=\pa_a -i\G_a$, produced 
in the course of the D-algebra. $z(\e)$ is a singular factor from momentum integration of divergent supergraphs and the 
$n$-dimensional $\hat{\d}_a^{~b}$ is produced from symmetric integration. Using the rules of dimensional reduction and the 
Bianchi identities in terms of covariant derivatives one can show that
\beq
{\rm{Tr} }\int d^4x~d^4\theta~\G^a\G_b(\d_a^{~b}-\hat{\d}_a^{~b})= -\e~{\rm{Tr} } \int d^4x~d^2\theta~
W^\a W_\a
\label{betag}
\eeq
From the above relation it is clear that in order to obtain a divergence the coefficient $z(\e)$ in (\ref{covstructure}) 
must contain at least a $1/\e^2$ pole. Moreover the complete result can be obtained by calculating tadpole-type contributions 
with a $\G^a\G_b \d_a^{~b}$ vertex and then covariantizing them by the substitution $\d_a^{~b}\rightarrow \d_a^{~b}-
\hat{\d}_a^{~b}$.
Thanks to all of this even the five loop computation becomes manageable.  

We describe here the main steps that apply both to the two-loop and to the five-loop calculation. As emphasized above we 
need consider graphs with internal chiral lines only.
Thus, according to the rules in \cite{GZ}, at a given order in loop one draws vacuum diagrams with chiral covariant propagators  
and $\Del^2$, $\bar{\Del}^2$ factors at the chiral vertices. 
Now, in order to reduce as much as possible the number of terms produced in the course of the $\Del$-algebra, we do not 
perform  the covariant $\Del$-integration at this stage but modify the procedure as follows.  We want to single out 
tadpole-type contributions proportional to $\G^a\G_a$, therefore we have to figure out which are the potential sources of 
such terms. 
The explicit representation of the chiral covariant propagators is given by
\beq
\Box_+=\frac{1}{2} \, \Del^a\Del_a-i W^\a \Del_\a -\frac{i}{2}( \Del^\a W_\a )\qquad \qquad
\Box_-= \frac{1}{2} \, \Del^a \Del_a-i \bar{W}^\ad \bar{ \Del}_\ad -\frac{i}{2}( \bar{\Del}^\ad \bar{W}_\ad)
 \label{covbox}
 \eeq
Therefore in the expressions above  we can disregard the terms involving the field strengths since they do not enter the 
structure in (\ref{covstructure}).
The $\G^a\G_a$ terms can arise only from the expansion of the covariant operator $\Del^a \Del_a$ or from contracted covariant 
derivatives $\Del^a\dots \Del_a$ produced while performing the $\Del$-algebra. The net result is that we can immediately 
expand the covariant propagators as follows
\beq
\frac{1}{\Box_{\pm}} \quad\rightarrow\quad \frac{1}{\frac{1}{2}\Del^a \Del_a}\quad \rightarrow \quad \frac{1}{\Box}~+~\frac{1}{2}~
\frac{1}{\Box} \G^a\G_a \frac{1}{\Box} 
\label{expprop}
\eeq
where $\Box=\frac{1}{2} \, \pa^a \pa_a$ is the flat propagator.
All the rest we drop since it will not contribute to the structure we are looking for.
In this way we obtain two types of diagrams: \\
I. the ones  with flat $D^2$ and $\bar{D}^2$ factors at the vertices, flat propagators and one $\G^a\G_a$ insertion, for 
which now standard D-algebra can be performed\\
and\\
II. the vacuum diagrams  with flat propagators but $\Del^2$, $\bar{\Del}^2$ factors at the chiral vertices in which the 
$\G^a\G_a$ vertex will have to appear after completion of the  $\Del$-algebra.\\
 The relevant terms will be the ones that after subtraction of ultraviolet and infrared subdivergences give rise to $1/\e^2$ 
contributions.
 
 At the two-loop level the analysis is very simple: the vacuum diagram to be considered is shown in Fig.2a. Following the 
procedure described above, it is straightforward to realize that only I-type diagrams can give rise to  $1/\e^2$ poles and 
so the calculation reduces to  the one presented in \cite{GZ}.

\begin{figure} [t]
\begin{center}
\begin{tabular}{cc}
\epsfysize=4cm\epsfbox{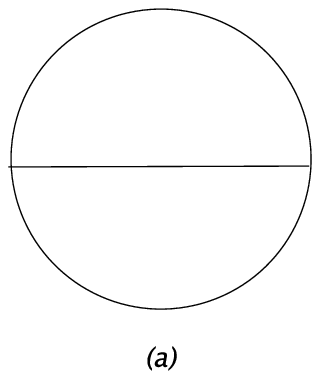} & \hspace{3cm} \epsfysize=4cm\epsfbox{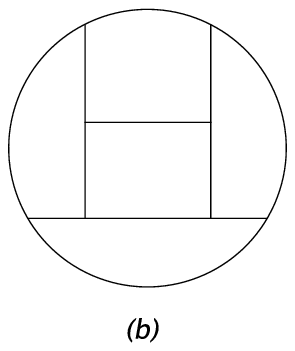}\\
\end{tabular}
\caption{Vacuum diagrams: (a) two-loops and (b) five-loops contributions}
\end{center}
\end{figure}

\begin{figure} [t]
\begin{center}
\begin{tabular}{c}
\epsfysize=4cm\epsfbox{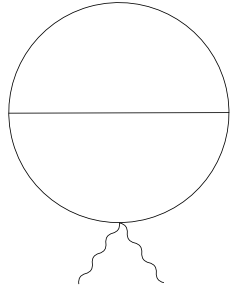}
\end{tabular}
\caption{Bosonic two-loop integral}
\end{center}
\end{figure}

We briefly summarize it here. Expanding the covariant propagators as in (\ref{expprop}) one obtains three times the 
diagram in Fig.3 which corresponds to the term
 \beq
 \frac{1}{2}~{\rm{Tr} }~(\G^a\G_a) \int \frac{d^n k ~d^n q}{(2\p)^{2n}} \frac{1}{q^2(q+k)^2 k^4}
 \label{2loopdiv}
 \eeq
 This integral contains a one-loop ultraviolet subdivergence and it is infrared divergent. It is convenient to remove the 
IR divergence using the $R^*$ subtraction procedure of \cite{CT}. After UV and IR subtraction one isolates the  $1/\e^2$ 
term and rewrites the result in a covariant form. Using (\ref{betag}) it can be recast in the standard divergent part of 
the two-loop effective action giving a total contribution
\beq
\frac{N}{(4\pi)^2}~\frac{3}{4}A~\frac{1}{\e}~{\rm{Tr} } 
\int d^4x~d^2\theta~W^\a W_\a
\label{betag2}
\eeq
where we have reinserted the $A$ factor defined in (\ref{AB}).

Now we are ready to attack the five-loop calculation which amounts to start with the vacuum diagram in Fig.2b. First we 
consider the I-type diagrams. In this case expanding the covariant propagators as in (\ref{expprop}) we produce twelve 
times the diagram in Fig.4. We perform standard D-algebra in the loops and look for a contribution that after subtraction 
of  IR and UV subdivergences gives rise to a $1/\e^2$ divergent term. One easily obtains a single contribution corresponding 
to the bosonic integral shown in Fig.4
\beq
\frac{1}{2}~{\rm{Tr} }~(\G^a\G_a) \int \frac{d^n k ~d^n q ~d^n r ~d^n s ~d^n t}
{(2\p)^{5n}} 
\frac{1}{r^2 (r+q)^2 s^2 (s+q)^2 t^2 (t+r)^2 (t+s)^2 (q+k)^2 k^4}
\label{5loopdiv}
\eeq
The IR divergence is treated as before via  $R^*$ subtraction \cite{CT} so 
that, inserting all the factors, the final result is given by
\beq
\frac{N}{(4\pi)^2} ~\frac{6}{5}B~\frac{1}{\e}~{\rm{Tr} } 
\int d^4x~d^2\theta~W^\a W_\a
\label{betag5}
\eeq
with $B$ defined in (\ref{AB}).

\begin{figure} [t]
\begin{center}
\begin{tabular}{ccc}
\epsfysize=4,5cm\epsfbox{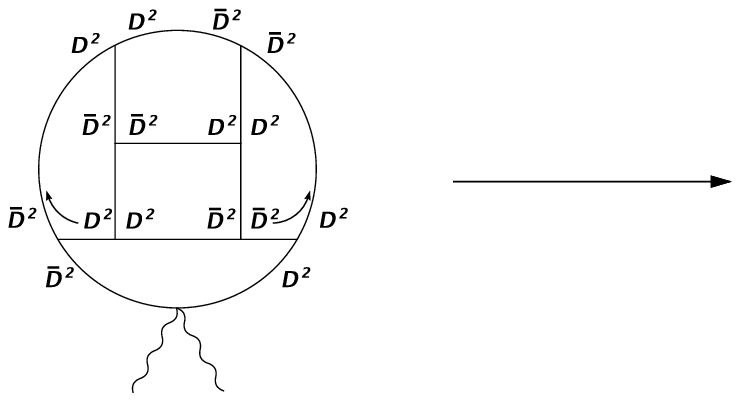}\hspace{0,7cm} &  \epsfysize=4,5cm\epsfbox{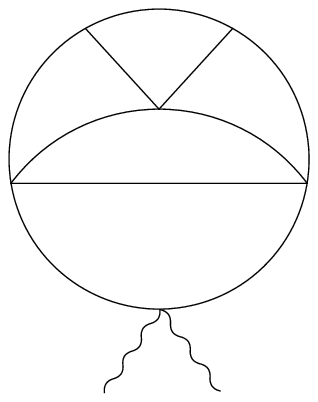}\\
\end{tabular}
\caption{ Five-loop supergraph and its associated relevant bosonic integral}
\end{center}
\end{figure}

In the class of II-type diagrams we have to analyze the vacuum diagram in Fig.5.  
We operate directly with the covariant spinor derivatives, pushing them through the propagators.
Unlike in ordinary D-algebra, covariant spinor derivatives and space-time derivatives contained in the propagators do not 
commute but it is easy to realize that they generate field strength factors which are not interesting for our calculation. 
Thus we can commute the $\Del_\a$'s through the $\Box^{-1}$.  The relevant contributions arise when we produce terms like
\bea
&&\Del^2 \bar{\Del}^2 \Del^2 =\Box_- \Del^2~~ \rightarrow ~~ -\frac{1}{2}~ \G^a\G_a \Del^2
\qquad\qquad \bar{\Del}^2 \Del^2  \bar{\Del}^2 =\Box_+ \bar{ \Del}^2~~ \rightarrow ~~ -\frac{1}{2}~ \G^a\G_a  \bar{\Del}^2 
\nonumber\\
&&~~~~~~~\nonumber\\
&& \Del_\a  \bar{\Del}_\ad  \Del^2 = i  \Del_a \Del^2 ~~\rightarrow ~~ \G_a  \Del^2 \qquad \qquad
\bar{\Del}_\ad  \Del_\a \bar{ \Del}^2 = i  \Del_a \bar{ \Del}^2 ~~\rightarrow ~~ \G_a \bar{ \Del}^2
\label{gammas}
\eea

\begin{figure} [ht]
\begin{center}
\begin{tabular}{ccc}
& \hspace{-0,3cm} \epsfysize=5,5cm\epsfbox{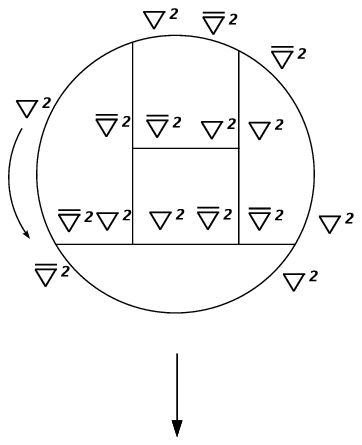} & \\ \\
  \epsfysize=4cm\epsfbox{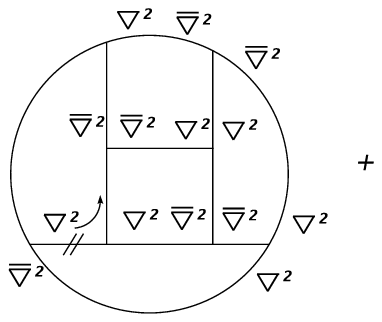} & \epsfysize=4cm\epsfbox{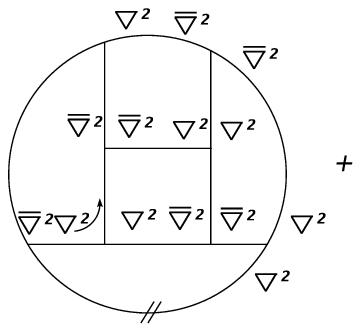}& \epsfysize=4cm\epsfbox{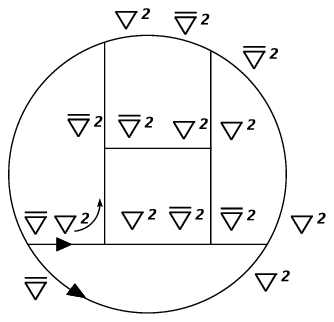}
\end{tabular}
\caption{Five-loop vacuum diagram and $\Del$-algebra operations}
\end{center}
\end{figure}

After all these preliminary observations, now one has to perform the covariant $\Del$-algebra explicitly and isolate the 
diagrams that could produce $1/\e^2$ ultraviolet divergences. It turns out that some cleverness must be used in order 
to reduce the number of the resulting contributions.
 We show in Fig.5 the successive manipulations that we used to obtain 
the final answer. As indicated in the figure the first integration by parts of the $\bar{\Del}^2$ factor produces three 
terms: we have denoted by 
\beq
/\!/ \, \equiv \frac{\frac{1}{2}\Del^a\Del_a}{\Box}~\rightarrow~1~-~\frac{1}{2}~ \frac{\G^a\G_a}{\Box}
\qquad \qquad \qquad \qquad   \blacktriangleright \,\, \equiv  \Del_a=\pa_a-i\G_a
\eeq

\begin{figure} [ht]
\begin{center}
\epsfysize=4cm\epsfbox{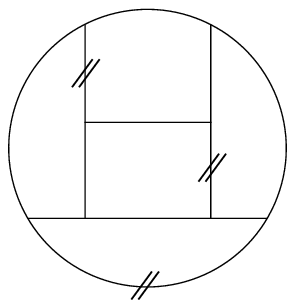}
\caption{Example of diagram not contributing to the $\frac{1}{\e^2}$ divergence}
\end{center}
\end{figure}
At this stage we have to work separately on the three graphs and complete the $\Del$-algebra by disregarding contributions
which do not contain  $1/\e^2$ divergent terms. 
(An example of diagram which is not interesting is the one shown in Fig.6. It
arises from the second diagram 
in Fig.5 and would produce only $1/\e$ divergent terms.) In fact if we move the $\Del$'s judiciously very few relevant terms are generated, the ones schematically 
shown in Fig.7. Now it is straightforward to show that by integration by parts these potentially relevant graphs do cancel 
out completely.

\begin{figure} [ht]
\begin{center}
\begin{tabular}{ccc}
\epsfysize=4cm\epsfbox{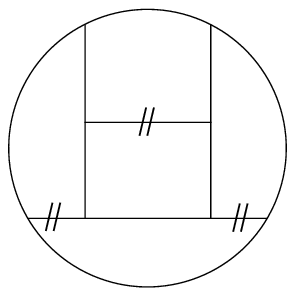} & \hspace{1,4cm}\epsfysize=4cm\epsfbox{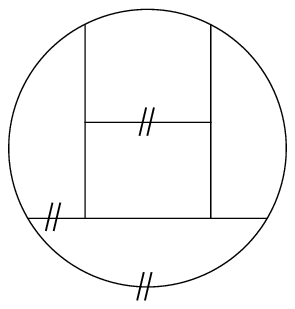} &\hspace{1,4cm} 
\epsfysize=4cm\epsfbox{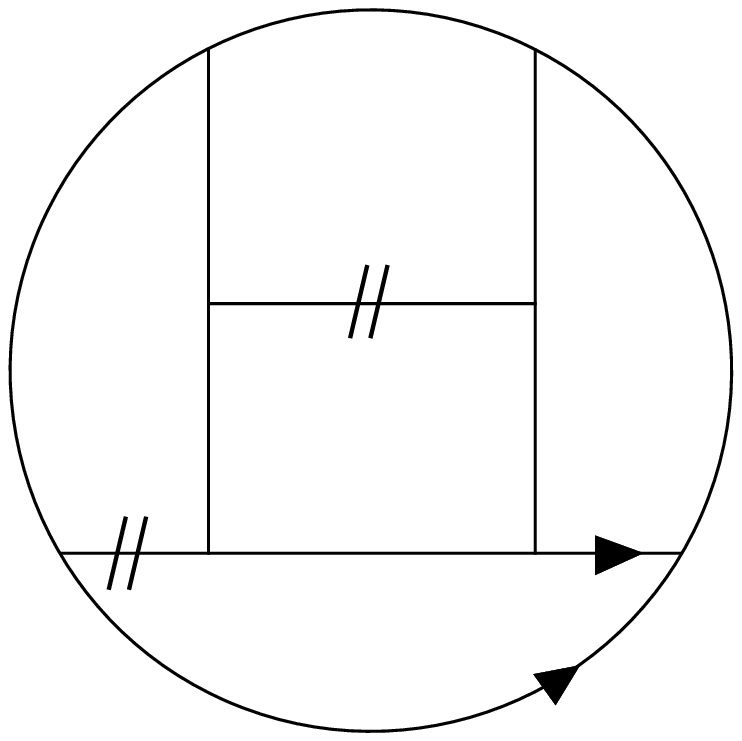} 
\end{tabular}
\caption{Relevant bosonic integrals associated to the five-loop graph of Fig.5}
\end{center}
\end{figure}
 
In conclusion, the only relevant contributions to the gauge beta function at order $g^{11}$ come from (\ref{betag2}) 
and (\ref{betag5}). Using the ordinary prescription to compute beta functions, we find
\beq
{\cal O}(g^{11}): \qquad\qquad \qquad \b_g=0 \qquad \Leftrightarrow \qquad A+4B=0
\eeq
Therefore a single condition on the $A$ and $B$ coefficients 
is sufficient to define the theory at its conformal point up to the order $g^8$
and to insure that, despite of the
non-finiteness of the theory, the gauge beta function vanishes at the next 
order. 

\vspace{2mm}

Now we want to come back to the fact that at order $g^8$ we have found that the theory subject to the condition in (\ref{gamma8}) is not finite. In order to proceed consistently we need renormalize the theory adding an appropriate counterterm.
As it follows from (\ref{forgamma}) this will be proportional to the divergence in the form
\beq
g^8 ~(A+B) ~(\frac{1}{\e} +\rho)
\label{counterterm}
\eeq
where $\rho$ is an arbitrary constant linked to the choice of a finite renormalization. It is worth noticing that the results obtained so far are completely independent of the subtraction scheme we have adopted. In fact even for the calculation of $\b_g$ at ${\cal{O}}(g^{11} )$ the arbitrary parameter $\rho$ does not enter in the evaluation of the coefficient of the $1/\e^2$ poles from which we read $\b_g$. The issue that now we want to address is what happens to the next order.

If we were to push the conformal invariance condition one order higher we should compute the chiral beta function 
at order $g^{10}$.  We have several sources of nontrivial contributions to $\g$ at this order:
one coming from the one-loop bubble proportional to $(a_5 + b_5)$, one from  two-loop diagrams  and one from five-loop diagrams.  In addition we need take into account the contribution from the counterterm in (\ref{counterterm}) which gives
\beq
g^{10} ~(A+B) ~(\frac{1}{\e} +\rho)~\frac{1}{\e} ~\left( \frac{\m^2}{p^2}\right)^{\e} 
\label{counter2}
\eeq
This last contribution is necessary to appropriately subtract diagrams that contain subdivergences at two and five loops, i.e. the ones that contain $1/\e^2$ poles considered in Section 2.
The condition for vanishing $\g$, obtained as usual from the finite $log$ terms,  gives an algebraic equation involving $A$, $B$ and $(a_5 + b_5)$ which, together with (\ref{gamma8})
allows to determine $A$ and $B$ parametrically and not necessarily vanishing. However the result depends unavoidably on the arbitrary constant  $\rho$ which appears in the form
\beq
(A+B) ~\rho
\label{scheme}
\eeq
If we wanted to kill the scheme dependence of the result we would need to impose $A+B=0$ which together with (\ref{gamma8}) would lead immediately to $A=B=0$, i.e. the theory is finite and  ${{\rm Im}\beta}=0$.
 
The comparison of these results with the ones of \cite{EMPSZ} as summarized in Section 2 leads to the conclusion that 
the request of conformal invariance via the vanishing of the beta functions is less restrictive than requiring
finiteness but the result is scheme dependent.

Pushing the calculations higher we expect to draw the same conclusion: conformal invariance via vanishing beta functions allows for   ${\rm Im}\beta\neq 0$ but this value and ultimately the conformal theory depend
on the choice of the particular renormalization scheme we use.

 \section{Differential renormalization approach}

In order to show that our findings do not depend on the particular regularization used in this Section 
we reconsider the calculation of the chiral propagator up to the order $g^8$ in the scheme of differential 
renormalization.

Differential renormalization works strictly in four dimensions. In its original formulation \cite{FJL} it is 
a renormalization without regularization, i.e. it allows for a direct computation of renormalized quantities
without the intermediate step of regularizing divergent integrals. In coordinate space the procedure consists   
in replacing locally singular functions (functions which do not admit a Fourier transform) with 
suitable distributions defined by differential operators acting on regular functions, where the derivatives
have to be understood in distributional sense.  
The simplest example is the function $1/(x^2)^2$ from the one--loop contribution to $\G^{(2)}$. This
function has a non-integrable singularity in $x = 0$. The prescription required by differential renormalization
in order to subtract such a singularity is

\begin{itemize}
\item
We substitute 
\beq
\frac{1}{x^4} ~\rightarrow~ -\frac{1}{4} \Box \frac{\log{M^2x^2}}{x^2}
\label{substitution}
\eeq
where $M$ is identified with the mass scale of the theory.

\item
We understand derivatives in the distributional sense, i.e.
\beq
\int d^4x f(x) \Box \frac{\log{M^2x^2}}{x^2} \equiv \int d^4x \Box f(x) \frac{\log{M^2x^2}}{x^2}
\eeq
for any regular function $f$.
\end{itemize}

The two expressions in (\ref{substitution}) are identical as long as $x \neq 0$, whereas they differ
by a singular term for $x \to 0$. The substitution (\ref{substitution}) can then be understood as adding 
a suitable counterterm \cite{FJMV}-\cite{S}:
\beq
\frac{1}{x^4} =  -\frac{1}{4} \Box \frac{\log{M^2x^2}}{x^2} ~+~ c(\a) \d^{(4)} (x)
\label{substitution2}
\eeq
where $c(\a)$ can be computed in some regularization scheme and becomes singular when the regularization
parameter $\a$ is removed.

Having in mind to study conformal invariance and/or finiteness for the deformed theory we need compute
both the renormalized chiral propagator and its divergent contributions. As long as we are interested in 
$\G_R^{(2)}$ we apply the standard differential renormalization prescription (\ref{substitution}) order by order
in $g^2$, whereas in order to identify the divergent counterterms which in (\ref{substitution}) are automatically
subtracted we need introduce a regularization prescription. We compute divergences using the analytic regularization \cite{BGG}.

As noticed above we are interested in computing the difference $(\G_{\rm deformed}^{(2)} - \G_{{\cal N}=4}^{(2)})$.
Thus at one-loop in coordinate space the contribution from the self-energy diagram is 
\bea
\label{oneloop}
\G^{(2)} &=& \frac{1}{x^4} ~ (h_1^2 + h_2^2 - 2g^2) \frac{N}{(4\pi^2)^2}
\\
&=& \frac{1}{x^4} ~ \left[ (a_1 + b_1 - 2)g^2 + (a_2 +b_2)g^4 + (a_3 +b_3)g^6 + (a_4 +b_4)g^8
+ \cdots \right]  \frac{N}{(4\pi^2)^2} \non
\eea
We renormalize this amplitude by the prescription (\ref{substitution}). At order $g^2$ we find the condition 
(\ref{order1}) which guarantees finiteness and vanishing of the beta functions.

As already discussed, once the condition (\ref{order1}) is satisfied we can neglect all higher loop diagrams 
which contain bubbles. In particular, at two and three loops we do not find relevant diagrams. Therefore, at orders 
$g^4$ and $g^6$ only the one-loop expression (\ref{oneloop}) contributes and
the conditions (\ref{order2-3}) are sufficient to cancel the renormalized and the divergent parts of  $1/x^4$.

At order $g^8$ the pattern changes since besides the contribution from 
(\ref{oneloop}) we have the new diagram 
in Fig.1. After D-algebra, in configuration space it corresponds to
\beq
-\frac{1}{2} (a_1 - b_1)^4 g^8 \frac{N^4}{(4\pi^2)^8} 
\, \frac{1}{x^2} \int \frac{d^4y \, d^4z \, d^4w}{y^2 z^2 (y-z)^2 (y-w)^2 (z-w)^2
(x-y)^2 (x-w)^2}
\label{4loopint}
\eeq
This expression has a singularity for $x \sim y \sim z \sim w \sim 0$. To compute its finite part, away from $x=0$ 
it is convenient to rescale the integration variables as $y \to |x| y$, $z \to |x|z$ and $w \to |x|w$. We are then
left with 
\beq
-\frac{1}{2} (a_1 - b_1)^4 g^8 \frac{N^4}{(4\pi^2)^8} 
\, \frac{1}{x^4} \int \frac{d^4y \, d^4z \, d^4w}{y^2 z^2 (y-z)^2 (y-w)^2 (z-w)^2
(1-y)^2 (1-w)^2}
\label{4loopint2}
\eeq 
The integral is finite and uniformly convergent for $x \to 0$. It has been
computed e.g. in \cite{Kaz} and it gives $20 \pi^6 \z(5)$.
At order $g^8$, summing this contribution to the one-loop result and renormalizing $1/x^4$ as in 
(\ref{substitution}) 
we obtain 
\beq
\G^{(2)}_R |_{g^8} =  (A + 4 B) \, \left( -\frac{1}{4\pi^2} \Box \frac{\log{M^2x^2}}{x^2} \right)
\label{finitecond}
\eeq
where $A$ and $B$ are given in (\ref{AB}). 

Therefore, the condition of vanishing $\g$ from $\G_R^{(2)}$  requires $A + 4B=0$. This is exactly the condition 
we have found working in dimensional regularization and momentum space. This is consistent 
with the fact that the Fourier transform of $\Box \frac{\log{M^2x^2}}{x^2}$ is $4\pi^2 \log{p^2}/{M^2}$.

Now we concentrate on the evaluation of the divergent contributions from the one-loop self-energy diagram and
from the four-loop diagram in Fig.1. Using analytic regularization in four dimensions, at one loop and order 
$g^8$ we have (for simplicity we neglect $(2\pi)$ factors)
\beq
A \frac{1}{(x^2)^{2+2\l}} 
\eeq
whereas at four loops we need evaluate the integral 
\bea
&& -\frac{N^4}{2} (a_1 - b_1)^4 g^8~\frac{1}{(x^2)^{1+\l}} \times 
\\
&& \int \frac{d^4y \, d^4z \, d^4w}{(y^2)^{1+\l} (z^2)^{1+\l} [(y-z)^2]^{1+\l} 
[(y-w)^2]^{1+\l} [(z-w)^2]^{1+\l} [(x-y)^2]^{1+\l} [(x-w)^2]^{1+\l}}
\non
\label{analreg}
\eea
Dimensional analysis allows to compute this integral and obtain
$(20 \z(5) + O(\l)) \frac{1}{(x^2)^{1+7\l}}$.
This gives the final answer $4B/(x^2)^{2+8\l}$ for the diagram in Fig.1. 

Now using the general identity
\beq
\frac{1}{(x^2)^{2+\a \l}} \sim  \frac{1}{\a \l} ~\d^{(4)}(x) ~+~ O(\l^0)
\eeq
and summing the one and four-loop results we find that the divergent contribution is
\beq
A \frac{1}{(x^2)^{2+2\l}} + 4B \frac{1}{(x^2)^{2+8\l}} ~\rightarrow~ (A+B)~\frac{1}{2\l} ~\d^{(4)}(x)
\eeq
Therefore the cancellation of divergences at order $g^8$ requires $A+B =0$.  If we were to compute the divergences 
arising at order $g^{10}$ we would find  results in total agreement
with the results  found using dimensional regularization.
Going higher in loops we would meet the same pattern an infinite number of times and we would be led to the final result  
for the coefficients as in (\ref{finalcoeff}).

\section{Conclusions}

We have reexamined the problem of finding superconformal fixed points
for $\b$--deformed SYM theories in the large $N$ limit and for the deformation parameter $\b$ generically 
complex. In a previous paper \cite{EMPSZ} we addressed this issue by requiring the theory to be finite. In this paper 
instead we have reformulated the problem by requiring the theory to
have vanishing beta functions.  

Looking for a surface of renormalization fixed points we have expressed the chiral couplings as power 
expansions in the gauge coupling $g$ (see eq. (\ref{expansion})). 
This introduces an infinite number of arbitrary coefficients which 
we fix by requiring order by order either finiteness or zero beta functions. 

This coupling constant reduction 
has an important consequence on the perturbative analysis of the theory. In fact we are forced to work 
pertubatively in powers of $g$ instead of powers of loops and at a given order different loops do mix. 
It follows that the condition of finiteness for the theory at a given order does not necessarily imply that the 
beta functions vanish and viceversa, in contrast with the case of a standard loop expansion.  

Collecting the present results and the ones in \cite{EMPSZ} the general situation can be then summarized as follows.
If we impose the cancellation of UV divergences at a given order we obtain conditions on the 
coefficients in the expansion (\ref{expansion}) which do not set automatically to zero the contribution
to the chiral beta function at the same order. In particular, in the planar limit the first nontrivial order 
where this happens is $g^8$. However, if we move one order higher and still require the cancellation of 
divergences we obtain more restrictive conditions on the coefficients and as a by--product 
all the beta functions at that order vanish. This pattern repeats itself at any order in pertubation theory
and leads to the following result: The finiteness condition selects a unique expansion (\ref{expansion}) for
$h_i(g)$ which corresponds to $\sinh{(2\pi {\rm Im}\b)} \sim (h_1^2 - h_2^2) = 0$, i.e. to a {\em real} deformation 
parameter $\b$. 

On the other hand, if we implement superconformal invariance by requiring directly vanishing  
beta functions regardless of finiteness we obtain less restrictive conditions on the coefficients in 
(\ref{expansion}) and more general solutions $h_i(g)$ to the renormalization group equation $F(g,h_i)=0$ which 
defines the surface of fixed points. These solutions correspond in general to theories which are not finite and
allow for a {\em complex} deformation parameter.

In our analysis the term ``finiteness'' is used in the standard way, that is to indicate a theory which does not
have UV divergences at any order in perturbation theory and, consequently, does not require any renormalization. 
In this sense finiteness is a well-defined and scheme independent statement. Its physical meaning is unquestionable 
since the set of couplings selected by this condition is uniquely fixed.
On the other hand, it is a matter of fact that in the presence of coupling constant reduction   
the conditions $\b_h, \b_g=0$ turn out to be scheme dependent. This means that the set of couplings which solve these
equations is not uniquely determined but depends on the renormalization scheme we chose. In particular, 
the generically complex value of the deformation parameter $\b$ that we find is scheme--dependent. 
This is the reason why in our approach finiteness and vanishing beta--functions are not equivalent statements.

A more general scenario can be obtained if we relax the request of scheme independence when imposing finiteness.
In dimensional regularization scheme dependence can be introduced by hands through the use of evanescent terms 
\cite{Kazakov87} in the reduction equations (\ref{expansion}). The extra freedom introduced by these 
$\e$--dependent terms allows to define the theory to be simultaneously finite and at its superconformal point for 
generically complex {\em but } scheme dependent $\b$ parameters, in agreement with \cite{EKT, Kazakov87, RSS2}. 
Therefore, the apparent discrepancy between our results and other statements in the literature 
\cite{EKT, Kazakov87, RSS2} can be traced back to the use of a different definition of finiteness.
    
In the presence of coupling constant reduction we are not guaranteed that finiteness theorems \cite{PW,GMZ} for the 
gauge beta function are true in their standard version. However, pushing the perturbative calculation up to
five loops, we have found that given the chiral beta function
vanishing at order $g^9$, then the gauge beta function is automatically zero at order $g^{11}$. Our result suggests 
that the finiteness theorems might be generalized as follows: If the matter chiral beta function vanishes up to the 
order $(g^{n})$ then the gauge beta function vanishes as well up to the order $(g^{n+2})$.

We have worked in the planar limit where the condition for superconformal invariance is known
exactly \cite{MPSZ}. However, the same pattern for finiteness vs. conformal 
invariance should appear also at finite $N$. This issue is presently under investigation.

\vspace{0.8cm}

\medskip

\section*{Acknowledgements}
\noindent 
This work has been supported in part by INFN, PRIN prot.2005024045-002
and the European Commission RTN program MRTN--CT--2004--005104.

\end{document}